\documentstyle[11pt,twoside]{article}
\leftline{\copyright~ 1999 International Press}
\leftline{Adv. Theor. Math. Phys. {\bf 2} (1998) {\ 1441 -- 1462} }

\begin{document}
\thispagestyle{empty}



\vspace{0.2in}

\begin{center}

{\huge \bf K\"ahler Cone Substructure }
\vspace{0.2in}

\renewcommand{\thefootnote}{}
\footnotetext{\small e-print archive: {\texttt http://xxx.lanl.gov/abs/hep-th/9810064}}

{\bf Eric Sharpe} \\
\vspace{0.2in}

Department of Physics \\
Box 90305 \\
Duke University \\
Durham, NC  27708 \\
{\tt ersharpe@cgtp.duke.edu} \\
\vspace{0.1in}

\end{center}
\begin{abstract}
To define a consistent perturbative geometric heterotic compactification
the bundle is required to satisfy a subtle constraint known
as ``stability,'' which depends upon the K\"ahler form.
This dependence upon the K\"ahler form is highly nontrivial -- 
the K\"ahler cone splits into subcones, with a distinct moduli space
of bundles in each subcone -- and has long been overlooked
by physicists.  In this article we describe this behavior
and its physical manifestation.
\end{abstract}



\section{Introduction}

To specify a perturbative heterotic compactification, one
must specify a bundle (or, more generally, a torsion-free sheaf)
on the compact space.
One cannot specify any bundle; rather, it must satisfy
certain consistency conditions in order to get a supersymmetric
low-energy theory.

One of the consistency conditions is that the bundle
must satisfy an equation known as the Donaldson-Uhlenbeck-Yau
equation.  This equation depends nontrivially upon the metric.
This metric dependence has long been ignored among string
theorists, but is in fact nontrivial and quite important.
\newpage 
\pagenumbering{arabic}
\setcounter{page}{1442}

\pagestyle{myheadings}
\markboth{\it K\"AHLER CONE SUBSTRUCTURE}{\it E. SHARPE}

In this paper we examine the metric dependence of this
consistency constraint on bundles in perturbative heterotic
compactifications.  In particular, on the Calabi-Yau manifolds
appearing, the K\"ahler cone splits into subcones (or ``chambers''), 
with a distinct
moduli space of bundles associated to each subcone.

We begin in sections~\ref{rrev} and \ref{mtstab} by reviewing 
general constraints on geometric
heterotic compactifications, and Mumford-Takemoto stability
in particular.  In section~\ref{kcsub} we then review the 
relevant mathematical results
concerning how the K\"ahler cone splits into subcones.
Then in section~\ref{phys} we examine the physical behavior
that reproduces this mathematics.  (Essentially, one gets
a perturbative enhanced $U(1)$ gauge symmetry on subcone walls, and the 
change in the moduli space is realized by examining D terms.)
In section~\ref{dual} we briefly consider the implications of
these results for string duality.  (For example, this behavior
in heterotic K3 compactifications corresponds to previously 
unknown behavior of hyper plets
in type IIA Calabi-Yau compactifications at intermediate
type IIA string coupling.)  Finally in section~\ref{cpx} we
work out a description of K\"ahler cone substructure on K3s that depends
only on the Riemannian metric, not the precise complex structure,
and also conjecture how this phenomenon generalizes when
the K\"ahler cone is complexified by adding a B field.
We also include a few appendices containing general background
on moduli space problems and certain technical derivations.

Although in principle similar remarks hold in general,
in this paper we will only consider bundles on surfaces
(and typically only on K3s), not in other dimensions.
We shall also only consider $GL(n,{\bf C})$ bundles, not bundles
with more general structure groups.  Finally, for most of
this paper we shall only explicitly refer to the classical K\"ahler cone,
not the complexified K\"ahler cone.  Only towards the end will we
explicitly study the effect of adding a B field.

In this article, when we speak of stability we shall always
be referring to Mumford-Takemoto stability.

\section{Rapid review of heterotic compactifications}   \label{rrev}

For a consistent perturbative compactification of either the
$E_8 \times E_8$ or $Spin(32)/{\bf Z}_2$ heterotic strings,
in addition to specifying a Calabi-Yau $Z$ one must also specify
a set of holomorphic vector bundles (or, more generally,
torsion-free sheaves\footnote{We mention sheaves for completeness,
though to aid readability in this paper we will only refer to bundles.})
$V_i$.  These vector bundles must obey two constraints.
For $GL(n,{\bf C})$ bundles one constraint can be written
as 
\begin{equation}    \label{duy1}
J^{n-1} \cdot c_1(V_i) \: = \: 0
\end{equation}
where $n$ is the complex dimension of the Calabi-Yau and $J$
is the K\"ahler form.  
Put another way, given a set of set of connections associated
to local coordinate trivializations that are Hermitian
(meaning, $F_{ij} = F_{\overline{\imath} \overline{\jmath}} = 0$),
the constraint can be written as 
\begin{equation}  \label{duy2}
g^{i \overline{\jmath}} F_{i \overline{\jmath}} \: = \: 0
\end{equation}
or, equivalently,
\begin{equation}   \label{duy3}
F \wedge J^{n-1} \: = \: 0
\end{equation}
This is known as the Donaldson-Uhlenbeck-Yau equation
\cite[section 15.6.2]{gsw}.

This constraint has a somewhat subtle
implication.  In general for any holomorphic bundle ${\cal E}$,
if there exists a Hermitian connection associated to ${\cal E}$ such
that, in every coordinate chart, the curvature $F$ satisfies
$F \wedge J^{n-1} \, = \, c I$, where $I$ is the identity
matrix and $c \in {\bf R}$ is a fixed chart-independent constant,
then ${\cal E}$ is either properly Mumford-Takemoto stable\footnote{We
shall explain stability momentarily.}, or Mumford-Takemoto semistable
and split \cite{kobayashi,donaldson,uhlenyau}.
Thus, the constraint in equation~(\ref{duy1}) implies that (but is not
equivalent to the statement) ${\cal E}$ is either stable,
or semistable and split.  In fact we can slightly simplify this
statement.  Properly semistable sheaves are grouped\footnote{More
precisely, points on a moduli space of sheaves that are properly semistable
do not necessarily correspond to unique semistable sheaves, but rather to 
``$S$-equivalence
classes'' of properly semistable sheaves.  Points that
are stable do correspond to unique stable sheaves -- $S$-equivalence
classes are a phenomenon arising only for properly semistable objects.
For more information, see appendix~\ref{modnotes}.}
in $S$-equivalence classes, and each $S$-equivalence class contains a unique
split representative \cite[p. 23]{huybrechtslehn}.

Thus, constraint~(\ref{duy1}) implies that ${\cal E}$ is Mumford-Takemoto
semistable.  Moreover, the constraint implies that the representative
of any $S$-equivalence class that is relevant for physics is
the unique split representative.

The other constraint is an anomaly-cancellation condition which, if
a single $GL(r,{\bf C})$ bundle $V_i$ is embedded in each $E_8$,
is often
written as
\begin{displaymath}
\sum_i \left( c_2(V_i) \, - \, \frac{1}{2} c_1(V_i)^2 \right)
\: = \: c_2(TZ)
\end{displaymath}
It was noted \cite{duffmined} that the anomaly-cancellation conditions
can be modified slightly by the presence of five-branes in the
heterotic compactification.  However, we shall only be
concerned with perturbative heterotic compactifications in this
paper.

\section{Mumford-Takemoto stability}  \label{mtstab}

In the previous section we mentioned that for a consistent
perturbative heterotic compactification, the holomorphic
bundle on the Calabi-Yau must be ``Mumford-Takemoto
semistable.''  What does this mean?

For any torsion-free sheaf ${\cal E}$, define the slope of 
${\cal E}$ to be
\begin{displaymath}
\mu({\cal E}) \: = \: \frac{ c_1({\cal E}) \cdot J^{n-1} }{ \mbox{rank }
{\cal E} }
\end{displaymath}
where $J$ is the K\"ahler form.  
We will sometimes use the notation $\mu_J({\cal E})$ when there is
ambiguity in the choice of K\"ahler form $J$.
We say ${\cal E}$ is Mumford-Takemoto
(semi)stable if, for all proper coherent subsheaves ${\cal F} \subset
{\cal E}$ such that $0 < \mbox{rank }{\cal F} < \mbox{rank }{\cal E}$
and ${\cal E}/{\cal F}$ is torsion-free, we have
\begin{displaymath}
\mu({\cal F}) \: ( \leq ) < \: \mu({\cal E})
\end{displaymath}

Note that if ${\cal E}$ is a torsion-free sheaf such that $c_1({\cal E})
= 0$ and if ${\cal E}$ has sections, then it can be at best semistable,
not strictly stable.  This is because the section defines a map
${\cal O} \rightarrow {\cal E}$, so we have a subsheaf ${\cal F}$
(namely, ${\cal F} = {\cal O}$) such that $\mu({\cal F}) = 0 =
\mu({\cal E})$.

In passing, we should make a technical remark concerning torsion-free
sheaves that are not bundles.  The stability constraint was
originally derived from the low-energy supergravity for geometric
compactifications involving bundles, not more general sheaves.
Although one can certainly define Mumford-Takemoto (semi)stability
for other sheaves (as we have done, in fact), there are other, 
inequivalent notions of stability (prominently, Gieseker stability),
and it is not clear whether Mumford-Takemoto stability is the
correct notion of stability for heterotic compactifications
involving sheaves that are not bundles.  For nongeometric 
compactifications, even less is known -- no one knows any 
analogue of the stability
constraint.

\section{K\"ahler cone substructure}  \label{kcsub}

Note that Mumford-Takemoto stability depends implicitly upon the
choice of K\"ahler form 
\cite{huybrechtslehn,friedbook,qin1,qin2,qin3,qin4,friedqin,matsukiwentworth,huli,gottsche,liqin,qinpriv}.
This choice is extremely important -- sheaves that are stable
with respect to one K\"ahler form may not be stable with
respect to another.  In general, for fixed Chern classes a moduli
space of sheaves will not have the same form everywhere inside
the K\"ahler cone, but rather will have walls along which
extra sheaves become semistable.  These walls stratify the
K\"ahler cone into subcones (or ``chambers''), inside any one of 
which the notion of stability is constant.

We should take a moment to clarify these remarks slightly.
In typical circumstances, a generic stable bundle will be stable
for all choices of K\"ahler form.  However, the stability of
some nongeneric subset will depend nontrivially upon the K\"ahler form,
and so the moduli space will change as one wanders around in the
K\"ahler cone.  More precisely, as the K\"ahler moduli are varied, some
(typically nongeneric) stable bundles will become strictly
semistable, then unstable, and vice-versa.

In this section we shall describe, without proof, necessary
(but not suffi-cient \footnote{More precisely, the conditions we shall
state are necessary (but in general not sufficient) for the moduli 
space to change.  
For rank greater than two, sufficient conditions on chamber walls for 
the moduli space to change are not known.}
if the rank is greater than 2) conditions for chamber walls inside the K\"ahler
cone.  (For the special case of moduli spaces of rank 2 sheaves,
it is known that these conditions are both necessary and sufficient
for the moduli space to change.)

Define the discriminant of a coherent sheaf ${\cal E}$ on an algebraic
K\"ahler surface to be
\begin{displaymath}
\Delta({\cal E}) \: = \: 2r c_2({\cal E}) \: - \: (r-1) c_1({\cal E})^2
\end{displaymath}
where $r$ is the rank of ${\cal E}$.  It can be shown
(see \cite[section 3.4]{huybrechtslehn} or \cite{friedbook,bog})
that when ${\cal E}$ is
Mumford-Takemoto semistable, $\Delta({\cal E}) \geq 0$.

Walls inside the K\"ahler cone $K$ are specified by divisors $\zeta$
satisfying certain conditions.  For a given divisor $\zeta$,
the corresponding wall is
\begin{displaymath}
W_{\zeta} \: = \: \left\{ J \in K \, | \, \zeta \cdot J = 0 \right\}
\end{displaymath}
(Note that by the Hodge index theorem, on an algebraic K\"ahler surface
the positive definite part of $H^{1,1}$ is one-dimensional,
so the intersection form on $H^{1,1}$ has signature
$(+,-,\cdots,-)$.)  

Precisely which divisors $\zeta$ can define chamber walls?
The conditions \cite{liqin,qinpriv} are that, for some integer $i$, 
$0 < i < r$, 
\begin{eqnarray*}
 & 1) & \zeta = r F - i c_1 \mbox{ for some divisor $F$} \\
 & 2) & -i (r-i) \Delta \leq \zeta^2 < 0
\end{eqnarray*}

We shall not attempt to completely prove this result here
(but for a detailed examination, see appendix~\ref{deriv}).
However,
part of this result is relatively clear.
Suppose ${\cal E}$ is a rank $r$ bundle that is stable in some parts of
the K\"ahler cone and unstable in others.
Let ${\cal F}$ be a potentially destabilizing subsheaf
of ${\cal E}$, in the sense that when ${\cal E}$ becomes
unstable as a function of the K\"ahler form $J$,
$\mu_J({\cal F})$ grows to become larger than $\mu_J({\cal E})$.  
Define $F = c_1({\cal F})$,
$i = \mbox{rank } {\cal F}$.
Then the condition on $J$ for the bundle ${\cal E}$ to be
strictly semistable is $\mu_J({\cal F}) = \mu_J({\cal E})$,
which we can rewrite as $J \cdot ( r F - i c_1 ) = 0$.
Put another way, if ${\cal E}$ is strictly semistable for some
K\"ahler form $J$
then there exists a divisor $F$ and an integer $i$,
such that $\zeta = r F - i c_1$ and $J \cdot \zeta = 0$.

Some examples might help the reader.  Consider a generic
elliptic K3 with section.  Its K\"ahler cone has two generators,
corresponding to the section $S$ and the fiber $F$, obeying
$S^2 = -2$, $F^2 = 0$, and $S \cdot F = 1$.
Write $J = a S + b F$, then the K\"ahler cone is defined by 
the inequalities $a > 0$ and $b > 2a$.
Consider a moduli space of rank 2 bundles of $c_1 = 0$ and
$c_2 = 4$.  In this case it is straightforward to show
that the K\"ahler cone splits into two subcones,
with the chamber wall located along K\"ahler forms proportional
to $S + 3 F$.  For another example, consider a moduli space consisting of
rank 2 bundles of $c_1 = 0$ and $c_2 = 24$ (the moduli space
containing the tangent bundle), on the same K3.  It is straightforward
to check that the K\"ahler cone splits into 15 subcones in this case.

In certain cases it is possible to see K\"ahler cone substructure
explicitly.  For example, in moduli spaces of equivariant sheaves
on toric varieties, this substructure is essentially manifest.
We shall not work through such examples in this paper;
see instead \cite{meallen,meallen2}.

Moduli spaces associated to distinct chambers of a K\"ahler
cone are often, but not always, birational to each other.
(If they are not birational, then it is because at least
one is reducible, and in crossing the wall an entire
component either appeared or disappeared.)
We shall see explicitly how distinct moduli spaces are related
in the next section.

In passing, we should note that the behavior of Mumford-Takemoto
stability as the K\"ahler form changes is closely related to the
behavior of GIT quotients under change of polarization
\cite{dolgachevhu,thaddeus}.

We should also point out that this behavior is closely related
to the behavior of Donaldson polynomial invariants on
manifolds of $b_2^+ = 1$ as the metric changes.  There also,
one finds walls in the space of metrics.  
(However, in Donaldson theory this only happens on four-manifolds
of $b_2^+ = 1$, whereas the K\"ahler cone substructure described
in this paper potentially occurs on any algebraic variety.)
In the present discussion
we fix Chern classes and examine the behavior of moduli spaces
of bundles of those fixed Chern classes as the K\"ahler form changes,
whereas in Donaldson theory one sums over contributions from
different Chern classes.  In essence, Donaldson theory is
a ``topological'' version of the ``algebro-geometric'' phenomenon
being discussed here.  For more information on Donaldson theory
on manifolds of $b_2^+ = 1$, see for example \cite{friedqin2,mooreed}.

\section{Physics at chamber walls}  \label{phys}

Suppose we have a heterotic compactification involving some
bundle ${\cal E}$, and we have varied the K\"ahler form until
the bundle ${\cal E}$ is no longer stable\footnote{In typical cases,
this can only happen for certain nongeneric ${\cal E}$.},
but rather properly
semistable.  What does this mean for the low-energy theory?
The low-energy theory picks up an enhanced $U(1)$ gauge symmetry.
The transformation of the moduli space of bundles is encoded
in D terms.

Why does the low-energy effective theory get an enhanced $U(1)$ gauge
symmetry?  Properly semistable bundles occur on a moduli space
in $S$-equivalence classes (see appendix~\ref{modnotes}),
and each $S$-equivalence class contains a unique split representative
\cite[p. 23]{huybrechtslehn}.
Moreover, as discussed earlier, 
the representative of the $S$-equivalence
class relevant for physics is the split representative.
Thus, if ${\cal E}$ is a properly semistable bundle, then the
physically relevant representative
of the same $S$-equivalence class can be written in the form
${\cal E} = {\cal F} \oplus {\cal G}$ for semistable ${\cal F}$, ${\cal G}$.
Since the bundle splits, we have a perturbative enhanced
$U(1)$ gauge symmetry.  This is because the low-energy gauge
theory is the largest subgroup of the ten-dimensional gauge group
that commutes with the structure group of the bundle.
When the bundle splits, its structure group can be reduced
from $SU(n)$ (where $n$ is its rank) to $S[ U(n_1) \times U(n_2)]$,
where $n = n_1 + n_2$.  Everything that commuted with 
$SU(n)$ also commutes with $S[U(n_1) \times U(n_2)]$,
and in addition there is an extra commuting $U(1)$
factor, described explicitly by $SU(n)$ matrices
of the form $\mbox{diag}(x, x, \cdots, x, y, y, \cdots y)$,
with $x^{n_1} y^{n_2} = 1$.
Thus, there is a perturbative enhanced
$U(1)$ gauge symmetry\footnote{In fact, we are being
slightly sloppy -- one sometimes will also need to mod out
by finite groups -- but this will not affect our
analysis.  It is relatively straightforward to see these
factors.  Suppose that we have embedded 
$SU(3)$ in an $E_8$.  The group $E_8$ contains a subgroup
\cite{allenpriv}
\begin{displaymath}
\frac{ E_6 \times SU(3) }{ {\bf Z}_3 }
\end{displaymath}
and so this is the reason why the largest group commuting
with $SU(3)$ is $E_6$ rather than $E_6 \times Z( SU(3) )$
($Z( SU(3) ) = {\bf Z}_3$) -- the center of $SU(3)$ is
identified with a ${\bf Z}_3$ subgroup of $E_6$.
If $SU(3)$ is reduced to $S[U(2) \times U(1)]$,
then the low-energy gauge group is actually
\begin{displaymath}
\frac{ E_6 \times U(1) }{ {\bf Z}_3 }
\end{displaymath}
We would like to thank A.~Knutson for an explanation of this
detail.}
in the low-energy effective theory.

Since the low-energy theory has picked up a $U(1)$, we now
have to worry about D terms, and indeed these will explicitly
realize the moduli space behavior mentioned earlier.  In order
to see this behavior explicitly, let us examine the chiral fields
of the low-energy theory which will be charged under the $U(1)$.

Write ${\cal E} = {\cal F} \oplus {\cal G}$, then deformations
of ${\cal E}$ (classified by elements of\footnote{Infinitesimal deformations
of an arbitrary torsion-free sheaf ${\cal E}$ are classified
by elements of $\mbox{Ext}^1({\cal E},{\cal E})$ \cite{meqik,ralph}, however
for the purposes of making this paper more readable, we usually
restrict to bundles, and for ${\cal E}$ a bundle, 
$\mbox{Ext}^1({\cal E},{\cal E}) = H^1({\it End } \, {\cal E})$.}
$H^1({\it End} \, {\cal E}) = 
H^1( {\it Hom}( {\cal E}, {\cal E}))$) have contributions from four sources:
\begin{displaymath}
\begin{array}{c}
H^1( {\it Hom}({\cal F}, {\cal F}) ) \\
H^1( {\it Hom}({\cal G}, {\cal G}) ) \\
H^1( {\it Hom}({\cal F}, {\cal G}) ) \\
H^1( {\it Hom}({\cal G}, {\cal F}) )
\end{array}
\end{displaymath}
The first two contributions -- namely, deformations of ${\cal F}$ and
${\cal G}$ individually -- are neutral under the $U(1)$.
The second pair of contributions, which mix ${\cal F}$ into 
${\cal G}$ and vice-versa -- have equal and opposite charges
under the $U(1)$.

In order to make notation more concise, let us define
\begin{eqnarray*}
\alpha_i & \in & H^1( {\it Hom}( {\cal F}, {\cal G}) ) \\
\beta_j & \in &  H^1( {\it Hom}( {\cal G}, {\cal F}) )
\end{eqnarray*}
and identify $\alpha_i$, $\beta_j$ with the corresponding
chiral superfields.

We can now write the D term\footnote{In this paper we concentrate
on K3 compactifications, and so we will get a triplet of D terms,
not a single D term.  However, to simplify the presentation,
we shall momentarily forget this point and only consider a single
D term.} of the low-energy effective action
associated to the enhanced $U(1)$ in the form
\begin{displaymath}
D \: = \: \sum_{i=1}^n | \alpha_i |^2 \: - \: \sum_{j=1}^m | \beta_j |^2
\: - \: r
\end{displaymath}
where $r$ is some function of the K\"ahler moduli, and
\begin{eqnarray*}
n & = & \mbox{dim }H^1({\it Hom}({\cal F}, {\cal G})) \\
m & = & \mbox{dim }H^1({\it Hom}({\cal G}, {\cal F})) 
\end{eqnarray*}
The analysis of these D terms is now essentially identical
to that presented in \cite{phases}.
Depending upon the values of $n$ and $m$, there are essentially
three distinct cases to consider:

1) $n = m = 0$

This trivial case occurs when locally all deformations of the
bundle preserve its splitness.  We shall not speak to this
case further. 

2) $n > 0$, $m = 0$

In this case, when $r > 0$ the moduli space is locally
${\bf P}^n$, yet when $r < 0$ the moduli space is empty.
This is precisely the description of a component of a reducible
moduli space appearing or disappearing at a chamber wall.

In the special case of bundles on K3s, however, $n = m$ by Serre duality,
so case (2) cannot arise.  (For a discussion of irreducibility
of moduli spaces of bundles on K3s, a sufficient but not necessary
condition for case (2) not to arise, see \cite{tomasphd,ogrady}.)

3) $n > 0$, $m > 0$

In this case, by varying $r$ we recover a birational transformation.

Recall from the previous section that moduli spaces
of bundles associated to distinct chambers are either
related by a birational transformation, or one is reducible
and an entire component appears or disappears at a chamber wall.
We have explicitly recovered these possibilities\footnote{We would
like to thank E. Witten for a useful discussion of this matter.}
in terms of D terms associated to the enhanced $U(1)$.

Now let us specialize to the case of a heterotic K3 compactification.
In this case in the low energy supergravity the moduli are
in hyperplets, and D terms come in triplets.
How does the above story specialize?
First, recall that a charged hyperplet can be thought of as
a pair of oppositely charged chiral plets.  In the present case,
using Serre duality we have
\begin{displaymath}
H^1({\it Hom}({\cal F}, {\cal G})) \: \cong \: H^1 ({\it Hom}({\cal G},
{\cal F}))^{\vee}
\end{displaymath}
so clearly the moduli that become charged under the $U(1)$ at the
subcone boundary fill out hyperplets.

We should clarify a few additional points.
A deformation by an element of $H^1({\it Hom}({\cal F},{\cal G}))$ or 
$H^1({\it Hom}({\cal G},{\cal F}))$ will define a distinct
bundle only up to an overall scale factor (see appendix~\ref{pext}), 
so we get a total
of 2 complex bosons (one hyperplet) more than we would have
gotten from $H^1({\it End } \, {\cal E})$ for generic
${\cal E}$.
Indeed, one expects\footnote{If the $U(1)$ were
nonanomalous, then this would follow immediately
from anomaly factorization -- if $n_H$ is the number
of hyper plets and $n_V$ the number of vectors,
then one needs $n_H - n_V = 244$ \cite{erler}.  However, we shall see shortly
that the $U(1)$ receives a mass through
the Dine-Seiberg-Witten mechanism, so an anomaly factorization
argument is not really appropriate.} light vectors to be paired with
light hyper plets, and as we have an enhanced $U(1)$,
we should not be surprised to find an extra hyper plet.

In addition, matter charged under the rest of the low-energy
gauge group can also become charged under the $U(1)$.
Recall from \cite{distlergreene} that $H^1({\cal E})$ counts
complex bosons charged under part of the (generic) low-energy
gauge group (phrased differently, $(1/2) h^1({\cal E})$ is the number of 
charged hyper plets).  In the case that ${\cal E} = {\cal F} \oplus {\cal G}$,
these will split into $h^1({\cal F})$ charged
complex scalars and $h^1({\cal G})$ charged complex scalars,
each charged oppositely under the enhanced $U(1)$.

In passing, we should also point out that in heterotic compactifications
on K3, D terms come in triplets, a fact we have essentially
ignored so far in this section.  One implication of this
fact is that K\"ahler cone substructure on K3s can be
understood in a complex-structure-invariant fashion;
this will be discussed in section~\ref{cpx}.

We have also glossed over anomaly cancellation so far 
in this section.  In fact, the enhanced $U(1)$ appearing
at chamber walls is anomalous, and gets a mass through
a six-dimensional version of the Dine-Seiberg-Witten
mechanism \cite{dinenatied}, as explained in \cite{douglasmoore}. 
The three real scalars forming the triplet of Fayet-Iliopoulos
terms in our discussion above form three-fourths of a hyper plet.
The fourth real scalar of this hyper plet gains a translation
symmetry, gauged by the $U(1)$, an artifact of Green-Schwarz
anomaly cancellation in ten dimensions.  (This fourth scalar
descends from the two-form tensor field in ten dimensions,
and its translation symmetry is a relic of the anomalous
transformation of the ten-dimensional tensor as assigned
in the Green-Schwarz mechanism.)  Because of this gauged
translation symmetry, the $U(1)$ vector field is actually
massive.  In particular, when checking six-dimensional Green-Schwarz
anomaly factorization, the $U(1)$ should not be counted.

\section{String duality}   \label{dual}

What can we learn by applying string duality to
K\"ahler cone substructure?

First, the attentive reader might be very concerned
about F-theory duals to heterotic K3 compactifications.
The K\"ahler cone substructure described in this paper
has not been seen in F-theory to date; why not?

Suppose we have compactified the heterotic theory on
an elliptic K3 with section.  Then, loosely speaking,
it turns out \cite{qin1} that there is a unique chamber
corresponding to the case that the section is much larger
than the fiber.  This particular chamber is the one
that has been sensed by F-theory arguments to date.

For example, in the recent past a description of
bundles on elliptic Calabi-Yaus was worked out by
Friedman, Morgan, Witten \cite{fmw}.  They explicitly
assume that the K\"ahler form lies in the distinguished
chamber described above, namely, that the elliptic
fiber is much smaller than all other K\"ahler moduli.
(Having made this assumption in the beginning, the
rest of their paper is written without any reference to
K\"ahler forms.)

Precisely what does the chamber structure dualize to?
For technical simplicity, consider a heterotic theory
compactified on $K3 \times T^2$, dual to type IIA on
some Calabi-Yau threefold.  Assume for simplicity that
the K3 is elliptic with section.  Then the weakly coupled IIA
string is dual to a heterotic theory with a big section
\cite{paul2}.  Put another way, the weakly coupled IIA
string is dual to the heterotic theory in the distinguished
chamber described above.  As we move to the boundary of
the chamber, the IIA string stops being weakly coupled.
Thus, it appears (somewhat loosely) that the chamber
structure of the perturbative heterotic string
dualizes to new behavior in the hyperplet moduli
space at intermediate IIA string coupling.

Since the IIA dual to K\"ahler cone substructure occurs
at intermediate string coupling, one probably cannot hope to
find a geometric explanation in terms of the IIA Calabi-Yau.
However, if one considered M theory compactifications
on a Calabi-Yau threefold, dual to the heterotic string
on $K3 \times S^1$, then it might be possible to see
heterotic K\"ahler cone substructure in terms of the
geometry of the Calabi-Yau.  We shall not pursue this
direction here, however.

\section{Complex structure ambiguity on K3s}   \label{cpx}

In most of this paper we have referred to bundles that are
holomorphic with respect to a specific complex structure.
On K3s we should really be slightly more careful.
As K3s are hyperK\"ahler manifolds, they possess Ricci-flat metrics
that are Hermitian with respect to multiple complex structures.
The choice of complex structure is not physically meaningful, and in 
heterotic compactifications on K3 this ambiguity is reflected
in the $SU(2)$ R-symmetry of the six dimensional supergravity.

As the low-energy supergravity possesses an $SU(2)$ R-symmetry that
reflects a physical invariance under rotation of complex structure,
one expects that there should exist a complex-structure-invariant
description of K\"ahler cone substructure.
More precisely, there should exist a description of K\"ahler cone
substructure that refers only to the metric and Chern classes,
not to any particular complex structure.

There is another reason to believe that there should exist
a complex-structure-invariant formulation of K\"ahler cone
substructure on K3s.  By following the arguments of
\cite[section 15.6]{gsw}, one finds that the metric and
connection must satisfy the Donaldson-Uhlenbeck-Yau 
equation~(\ref{duy2}) for not just one complex structure,
but for each complex structure on the K3.  Put another
way, one has a triplet of constraints of the form
\begin{displaymath}
F \wedge J^{ij} \: = \: 0
\end{displaymath}
where $J^{ij}$ is a triplet of K\"ahler forms, transforming
under the symmetric representation of $SU(2)_R$.

Such a complex-structure-invariant description does exist, 
and we shall describe it here,
following the notation and conventions of \cite{pauldave,paulrev}.

First, we shall consider how K\"ahler cone substructure is specified
on the space of Ricci-flat metrics on K3, namely the
Grassmannian of spacelike $3$-planes in ${\bf R}^{3,19}$.
Suppose we are studying moduli spaces of bundles of rank $r$
and Chern classes $c_1$, $c_2$.  We shall make the (simplifying)
assumption that $c_1 = 0$.  (Note that if $c_1$ were nonzero,
then it would implicitly fix a complex structure, namely
that in which $c_1 \in H^2(K3, {\bf Z}) \cap H^{1,1}(K3, {\bf R})$.)
Let $\zeta \in \Gamma^{3,19}$ such that
for some integer $i$, $0 < i < r$, 
\begin{displaymath}
-i \, (r-i) \, \left( 2r c_2  \right) \: \leq \:
r^2 \zeta^2 \: < \: 0
\end{displaymath}
then in the space of Riemannian metrics (defined by spacelike 
$3$-planes $\Sigma \subset \Gamma^{3,19}$)
there is a wall defined by $\zeta$, as
\begin{displaymath}
W_{\zeta} \: = \: \left\{ \Sigma \in \mbox{Grassmannian(spacelike $3$-planes
in ${\bf R}^{3,19}$)} \, | \, \Sigma \cdot \zeta \: = \: 0 \, \right\}
\end{displaymath}
and along this wall, for certain nongeneric bundles,
one will get enhanced $U(1)$ gauge symmetries in the low-energy
theory.

Specific complex structures are then assigned to a Riemannian 
metric $\Sigma$ by 
picking a
specific spacelike $2$-plane $\Omega \subset \Sigma$,
and the K\"ahler form $J$, modulo rescalings, is simply the 
orthogonal complement of $\Omega$ in $\Sigma$.
For any given choice of $\Omega$, this description correctly
reproduces the K\"ahler cone substructure described earlier, and so
this is clearly the desired complex-structure-invariant formulation.

If one is somewhat braver, one can make a natural conjecture concerning
how this situation changes when one adds a $B$ field.
In this case, let $\zeta \in \Gamma^{4,20}$ be such that, for some
integer $i$, $0 < i < r$,
\begin{displaymath}
-i \, (r-i) \, \left( 2r c_2 \right) \: \leq \:
r^2 \zeta^2 \: < \: 0
\end{displaymath}
Metrics and $B$ fields are combined into a spacelike $4$-plane
$\Pi \subset {\bf R}^{4,20}$.  It then seems quite
reasonable that walls in the space of
spacelike $4$-planes are defined by
\begin{displaymath}
W_{\zeta} \: = \: \left\{ \Pi \in \mbox{Grassmannian(spacelike $4$-planes
in ${\bf R}^{4,20}$)} \, | \, \Pi \cdot \zeta \: = \: 0 \, \right\}
\end{displaymath}
How do we then break down $\Pi$ into a specific choice of metric
and $B$ field ?
Pick $w \in \Gamma^{4,20}$ such that $w^2 = 0$ and $w \cdot \zeta = 0$.
Define $\Sigma' = \Pi \cap w^{\perp}$, then a metric is defined
by a spacelike $3$-plane $\Sigma$ which is simply the image of
$\Sigma'$ in $w^{\perp} / w$.
In order to find a corresponding value of $B$,
pick $w^* \in \Gamma^{4,20}$ such that $(w^*)^2 = 0$, $w \cdot w^* = 1$,
$w^* \perp w^{\perp}$, and $w^* \cdot \zeta = 0$.
Let $B' \in {\bf R}^{4,20}$ such that $B' \cdot \Sigma' = 0$
and $B' \cdot w = 1$,
then project $B'$ into $w^{\perp}/w$ to get $B$.

Note that this condition on the conformal field theory
is very closely related to the condition for an enhanced
nonabelian gauge symmetry in type IIA compactifications on K3
\cite{paulrev}.

\section{Conclusions}

In this paper we have studied the ``stability'' of bundles,
a necessary condition for a consistent perturbative heterotic
compactification, and in particular examined the dependence
of stability upon the K\"ahler form.  The K\"ahler cone splits into
subcones, which is reflected physically in the appearance
of an enhanced $U(1)$ gauge symmetry at a subcone wall (for
a compactification involving a (nongeneric) bundle whose
stability changes at the wall).  In particular, we have outlined
known mathematical results concerning the positions of these
walls, and also examined how their appearance is reflected
physically. 

There are several directions that remain to be pursued.
For example, we have not studied K\"ahler cone substructure
on Calabi-Yaus of dimension greater than two.  To our knowledge,
mathematical results on stability for algebraic varieties of dimension
greater than two are extremely limited (though it may be possible
to empirically derive some conjectures by using equivariant
sheaves \cite{meallen,meallen2} on toric varieties, for which K\"ahler cone
substructure appears manifestly.)

We also have not studied bundles with structure group other
than \newline $GL(n,{\bf C})$.  Presumably related phenomena appear in spaces
of bundles of other structure groups; it would be interesting
to determine precisely what phenomena occur.

\section{Acknowledgements}

We would like to thank P.~Aspinwall, R.~Friedman, T.~Gomez, A.~Knutson,
G.~Moore, 
J.~Morgan, 
D.~Morrison, R.~Plesser, Z.~Qin, and E.~Witten for useful discussions.

\appendix

\section{Notes on moduli spaces of bundles}    \label{modnotes}

In order to get a well-behaved
moduli space, one needs some notion of ``stability.''
For example, consider the analogous (and much simpler) problem of
constructing ${\bf P}^1$ as a quotient
of ${\bf C}^2$ by ${\bf C}^{\times}$:
\begin{displaymath}
{\bf P}^1 \: = \: \frac{ {\bf C}^2 \, - \, \{ 0 \} } { {\bf C}^{\times} }
\end{displaymath}
In order to get a well-behaved result, one first removes
the point $0$ from ${\bf C}^2$ before quotienting by ${\bf C}^{\times}$.
Technically this construction of ${\bf P}^1$ is known as
a Geometric Invariant Theory (GIT) quotient, and the point
$0 \in {\bf C}^2$ is an example of an unstable point.

Alternatively, one can construct ${\bf P}^1$ as
\begin{displaymath}
{\bf P}^1 \: = \: \frac{ S^3 }{ U(1) }
\end{displaymath}
In this construction, we do not need to remove a point before
quotienting; rather, we first restrict to a cross-section of
${\bf C}^2$, and then quotient by a smaller group.
This construction is known formally as a symplectic quotient,
and as is well-known, GIT quotients give the same results
as symplectic quotients.  In symplectic quotients
one does not define a notion of ``stability'' -- one picks
out a slice at the beginning that never intersects any ``unstable''
points.

Strictly speaking string theory
uses a description of moduli spaces closely analogous to
a symplectic quotient -- solutions of the Donaldson-Uhlenbeck-Yau
equation.  However, this analogue of a symplectic quotient
has an equivalent formulation in terms of an analogue of
a GIT quotient, and it is the GIT quotient formulation that
most people refer to.

So far we have considered the gross features of moduli spaces,
however for this paper we shall need a little more information.
It is possible for a point to be stable, but only just barely
 -- this is often referred to as being semistable.
Semistable objects do not map one-to-one into the moduli space;
rather, to get well-behaved results, several semistable
objects are typically identified with a single point on
the moduli space.  In the context of moduli spaces of
bundles, such classes of semistable holomorphic bundles
are known as $S$-equivalence classes. 
(In the context of symplectic quotients, this corresponds to
a symplectic reduction at a nonregular value of the moment map.)

Now let us see how $S$-equivalence classes arise a little more
explicitly.
Consider deforming ${\cal E} = {\cal F} \oplus {\cal G}$
by an element of $H^1({\it Hom}({\cal F}, {\cal G}))$ only -- 
do not turn on any elements of $H^1({\it Hom}({\cal G}, {\cal F}))$.
Then it turns out (see appendix~\ref{pext}) that the overall
scale does not matter -- elements of $H^1({\it Hom}({\cal F},{\cal G}))$
which differ only by a scale define isomorphic bundles.

Now, by scaling an element of $H^1({\it Hom}({\cal F},{\cal G}))$ down
to zero, we recover a one-parameter family of bundles, of the form
\begin{displaymath}
{\cal E}_t \: = \: \left\{ \begin{array}{ll}
        {\cal E}' & t \neq 0 \\
        {\cal F} \oplus {\cal G} & t = 0
        \end{array} \right.
\end{displaymath}
If we define a moduli space of bundles in such a way that
${\cal E}'$ and ${\cal E}_0 = {\cal F} \oplus {\cal G}$ are
distinct, then as ${\cal E}_0$ is the limit of a one-parameter
family describing ${\cal E}'$, the moduli space cannot
be Hausdorff\footnote{Technical experts would no doubt prefer
we use the term ``separable'' rather than ``Hausdorff,''
however we have decided to forgo a small amount of technical
accuracy in exchange for improved readability.}.  It is therefore
much more natural to associate ${\cal E}'$ and ${\cal E}_0$
with the same point on the moduli space, thereby getting a 
much better behaved moduli space.  The pair consisting of
${\cal E}'$ and ${\cal E}_0$ is an example of an $S$-equivalence
class, and this example has hopefully demonstrated why 
one uses $S$-equivalence classes to define moduli spaces.

In the text, we have discussed how stability depends upon
the K\"ahler form.  What happens to $S$-equivalence
classes as the K\"ahler form is varied?  Let ${\cal E}$ be a 
bundle that is stable
with respect to some K\"ahler forms and unstable with respect to 
others, so that along some wall in the K\"ahler cone it is
strictly semistable.  As we have just discussed, along
that wall in the K\"ahler cone, ${\cal E}$ will
lie in a nontrivial $S$-equivalence class of bundles.
What typically happens is that the other elements of the $S$-equivalence
class are strictly semistable only for K\"ahler forms along that
wall, and are unstable for other K\"ahler forms.  So, if the
K\"ahler form is moved off the wall in such a 
way that ${\cal E}$ becomes stable, then the other elements of
the $S$-equivalence class become unstable and so drop out
of the picture, leaving one with a unique representative.

\section{Derivation of chamber walls}   \label{deriv}

In this section we shall derive the necessity conditions
for a chamber wall in the K\"ahler cone, in a special case.
This derivation is already known (see for example \cite{liqin});
however, as the methods are unfamiliar to most physicists, we repeat
it here.

Before working through the derivation, we need a few standard
results.  Given a torsion-free coherent sheaf, there exists
a filtration of ${\cal E}$ \cite{huybrechtslehn,friedbook}
\begin{displaymath}
0 = {\cal E}_0 \: \subset \: {\cal E}_1 \: \subset \: \cdots \:
\subset {\cal E}_n = {\cal E}
\end{displaymath}
(where $\subset$ indicates proper subsheaf)
known as the Harder-Narasimhan filtration, 
with the properties
\begin{eqnarray*}
& (1) & {\cal E}_i / {\cal E}_{i-1} \mbox{ is semistable for } 1 \leq 
i \leq n \\
& (2) & \mu({\cal E}_i / {\cal E}_{i-1}) > \mu( {\cal E}_{i+1} / {\cal E}_i )
\mbox{ for } 1 \leq i \leq n-1 
\end{eqnarray*}
For every torsion-free coherent sheaf ${\cal E}$, a Harder-Narasimhan
filtration exists and, for fixed K\"ahler form, is unique.

Note that the Harder-Narasimhan filtration is trivial (meaning, of
the form $0 = {\cal E}_0 \subset {\cal E}_1 = {\cal E}$) precisely
when ${\cal E}$ is semistable.  Intuitively, the Harder-Narasimhan
filtration gives information about the ``instability'' of ${\cal E}$.

Now, suppose ${\cal E}$ is a torsion-free coherent sheaf
on an algebraic surface that is stable with respect to K\"ahler form
$J_1$ and unstable with respect to K\"ahler form $J_2$.
Suppose ${\cal E}$ has Harder-Narasimhan filtration (with respect
to $J_2$) of the form
\begin{displaymath}
 0 \: \subset \: {\cal E}_1 \: \subset \: {\cal E}
\end{displaymath}
This need not always be the case; however, 
we shall only study this case in this appendix.
(The general case follows similarly \cite{liqin,qinpriv}.)
We have an exact sequence
\begin{displaymath}
0 \: \rightarrow \: {\cal E}_1 \: \rightarrow \: {\cal E} \:
\rightarrow \: {\cal E} / {\cal E}_1 \: \rightarrow \: 0
\end{displaymath}

Define the discriminant $\Delta({\cal E}) = 2r c_2({\cal E}) - 
(r-1) c_1({\cal E})^2$, where $r = \mbox{rank }{\cal E}$.
By Bogomolov's inequality \cite{huybrechtslehn,friedbook,bog},
we have that $\Delta({\cal E}) \geq 0$ if ${\cal E}$ is
Mumford-Takemoto semistable.

Define $F = c_1({\cal E}_1)$, $i = \mbox{rank }{\cal E}_1$,
and $\zeta = r F - i c_1({\cal E})$.
Use the identity
\begin{displaymath}
\Delta({\cal E}) \: - \: \frac{r}{i} \Delta({\cal E}_1)
\: - \: \frac{r}{r-i} \Delta({\cal E}/{\cal E}_1) \: = \:
- \frac{\zeta^2}{i(r-i)}
\end{displaymath}
Now, as ${\cal E}_1$ and ${\cal E}/{\cal E}_1$ are $J_2$-semistable
(from the definition of Harder-Narasimhan filtration),
we have that
\begin{displaymath}
\Delta({\cal E}) \: \geq \: - \frac{ \zeta^2 }{ i (r-i) }
\end{displaymath}

We now merely need to show that $\zeta^2 < 0$.
From the definition of Harder-Narasimhan filtration,
we have $\mu_{J_2}({\cal E}_1) > \mu_{J_2}({\cal E}/{\cal E}_1)$,
which implies $J_2 \cdot \zeta > 0$.
As ${\cal E}$ is $J_1$-stable, we know that
$\mu_{J_1}({\cal E}_1) < \mu_{J_1}({\cal E})$,
which implies that $J_1 \cdot \zeta < 0$.
Thus, there exists $J$ such that $J \cdot \zeta = 0$,
and so from the Hodge index theorem we have $\zeta^2 < 0$.

Thus, if ${\cal E}$ is $J_1$-stable but $J_2$-unstable,
then there exists a divisor $\zeta$ such that, for some
integer $i$, $0 < i < r$,
\begin{eqnarray*}
& (1) & \zeta = r F - i c_1({\cal E}) \mbox{ for some divisor } F \\
& (2) & -i (r-i) \Delta({\cal E}) \leq \zeta^2 < 0
\end{eqnarray*}
and moreover ${\cal E}$ becomes strictly semistable for
a K\"ahler form $J$ such that $J \cdot \zeta = 0$.
It should be clear that these conditions are necessary
for a moduli space to change, but not necessarily sufficient.

\section{Notes on $H^1({\it Hom}({\cal F},{\cal G}))$}  \label{pext}

Suppose ${\cal F}$ and ${\cal G}$ are both bundles.
Then we can define a bundle ${\cal E}$ as an extension
of ${\cal F}$ by ${\cal G}$
\begin{displaymath}
0 \: \rightarrow \: {\cal G} \: \rightarrow \: {\cal E} \:
\rightarrow \: {\cal F} \: \rightarrow \: 0
\end{displaymath}
and extensions of this form are well-known to be classified
by elements of $\mbox{Ext}^1({\cal F},{\cal G}) = H^1({\it Hom}({\cal F},
{\cal G}))$.

There is, however, a subtlety:  distinct extensions are not necessarily
the same thing as distinct bundles.
In fact, elements of $H^1({\it Hom}({\cal F},{\cal G}))$ which
differ only by an overall scale define isomorphic bundles.

Thus, isomorphism classes of non-split bundles are actually
classified by ${\bf P} \mbox{Ext}^1({\cal F},{\cal G}) = 
{\bf P} H^1({\it Hom}({\cal F},{\cal G}))$.

How can we see this fact explicitly?
Construct the bundles ${\cal F}$, ${\cal G}$, and ${\cal E}$
in terms of transition functions on overlaps of coordinate charts.
The transition functions for ${\cal E}$ can be written in the
form
\begin{displaymath}
\left[ \begin{array}{cc}
       *_{\cal F} & A \\
       0 & *_{\cal G} 
       \end{array}    \right]
\end{displaymath}
where $*_{\cal F}$ and $*_{\cal G}$ are transition functions for
${\cal F}$, ${\cal G}$, respectively, and the $A$'s on each overlap
determine an element of $H^1({\it Hom}({\cal F}, {\cal G}))$.
Note that if $A = 0$ on each coordinate overlap, then 
they define the zero element of $H^1({\it Hom}({\cal F},{\cal G}))$,
and in particular the bundle ${\cal E}$ splits:  ${\cal E} = {\cal F}
\oplus {\cal G}$.

Suppose we multiply the element of $H^1({\it Hom}({\cal F},{\cal G}))$
by a scalar $t$, then the effect is to multiply the $A$ in each
transition function by $t$.  We claim that the resulting bundle, call it
${\cal E}_t$, is isomorphic to the original bundle ${\cal E}$.
To see this, simply change the local trivializations by multiplication
by 
\begin{displaymath}
\left[ \begin{array}{cc}
       1 & 0 \\
       0 & t 
       \end{array} \right]
\end{displaymath}
then the transition functions for ${\cal E}_t$ become those of ${\cal E}$:
\begin{displaymath}
\left[ \begin{array}{cc}
       1 & 0 \\
       0 & t 
       \end{array} \right]
\, 
\left[ \begin{array}{cc}
       *_{\cal F} & tA \\
       0 & *_{\cal G}
       \end{array} \right]
\,
\left[ \begin{array}{cc}
       1 & 0 \\
       0 & t^{-1} 
       \end{array} \right]
\: = \:
\left[ \begin{array}{cc}
       *_{\cal F} & A \\
        0 & *_{\cal G}
       \end{array} \right]
\end{displaymath}
In other words, ${\cal E}_t \cong {\cal E}$ for $t \neq 0$, 
precisely as claimed.

Similar results hold when ${\cal F}$ and ${\cal G}$ are more
general coherent sheaves, namely, elements of $\mbox{Ext}^1({\cal F},
{\cal G})$ that differ only by an overall scale define
isomorphic sheaves.  The argument is somewhat different
than was used above in the special case of bundles;
for completeness, we outline it here\footnote{We would like to
thank T. Gomez for a useful discussion of this matter.}.

First, consider an injective resolution of ${\cal G}$:
\begin{displaymath}
0 \: \longrightarrow \: {\cal G} \: \longrightarrow \:
{\cal I}_0 \: \stackrel{ d_0 }{\longrightarrow} \: {\cal I}_1 \: 
\stackrel{ d_1 }{\longrightarrow} \:
\cdots 
\end{displaymath}
Now, calculate $\mbox{Ext}^1({\cal F},{\cal G})$ as
a right derived functor of $\mbox{Hom}({\cal F},-)$,
namely, as the cohomology of the complex
\begin{displaymath}
\mbox{Hom}({\cal F}, {\cal I}_0) \: \stackrel{ d'_0 }{\longrightarrow}
\: \mbox{Hom}({\cal F},{\cal I}_1) \: \stackrel{ d'_1 }{\longrightarrow}
\: \mbox{Hom}({\cal F},{\cal I}_2) \: \stackrel{ d'_2 }{\longrightarrow}
\: \cdots
\end{displaymath}
In other words, represent an element of $\mbox{Ext}^1({\cal F},{\cal G})$
by an element $h \in \mbox{Hom}({\cal F},{\cal I}_1)$
such that $d'_1(h) = 0$, modulo $\mbox{im } d'_0$.

More concretely, given $h \in \mbox{Hom}({\cal F},{\cal I}_1)$ such
that $d'_1(h) = 0$, we have the following
diagram
\begin{displaymath}
\begin{array}{ccccccccc}
0 & \longrightarrow & {\cal G} & \longrightarrow & {\cal E} 
& \longrightarrow & {\cal F}
& \longrightarrow & 0 \\
 & & \| & & \downarrow & & \downarrow h & & \\
 0 & \longrightarrow & {\cal G} & \longrightarrow & {\cal I}_0 &
\stackrel{ d_0 }{\longrightarrow} & {\cal I}_1 & \longrightarrow &
\cdots 
\end{array}
\end{displaymath}
where the extension ${\cal E}$ is defined as follows.
Local sections of ${\cal E}$ are given by pairs $(f,i)$,
where $f$ is a local section of ${\cal F}$ and $i$ is a local
section of ${\cal I}_0$ obeying the constraint $h(f) = d_0(i)$.
The maps ${\cal E} \rightarrow {\cal F}$ and
${\cal E} \rightarrow {\cal I}_0$ are the obvious projections.
Verification that the diagram above is commutative and of related 
details is left as an exercise for the reader.

We can now check that multiplying $h$ by a scalar $t$
yields an isomorphic sheaf.  Define $h' = t h$, and
${\cal E}_t$ the new extension.  If $(f,i)$ is a local
section of ${\cal E}$, then $(f/t,i)$ is a local
section of ${\cal E}_t$, because $h(f) = h'(f/t)$.

Thus for $t \neq 0$, ${\cal E}$ and ${\cal E}_t$ are isomorphic as sheaves,
the isomorphism sending $(f,i)$ to $(f/t,i)$.

Finally, we should perhaps clarify why two sheaves that are
isomorphic as sheaves need not be isomorphic as extensions.
For two sheaves ${\cal E}_1$ and ${\cal E}_2$ to be
isomorphic as extensions of ${\cal F}$ by ${\cal G}$ means
that the following diagram must commute:
\begin{displaymath}
\begin{array}{ccccccccc}
0 & \longrightarrow & {\cal G} & \longrightarrow & {\cal E}_1
& \longrightarrow & {\cal F} & \longrightarrow & 0 \\
 & & \| & & \downarrow p & & \| & & \\
0 & \longrightarrow & {\cal G} & \longrightarrow & {\cal E}_2
& \longrightarrow & {\cal F} & \longrightarrow & 0 
\end{array}
\end{displaymath}
where $p: {\cal E}_1 \rightarrow {\cal E}_2$ is the isomorphism
in question.  It is easy to check that the isomorphism given
above between ${\cal E}$ and ${\cal E}_t$ as sheaves does not
yield a commutative diagram of the above form,
thus they cannot be isomorphic as extensions.

\end{document}